\documentclass[12pt,preprint]{aastex}

\usepackage{epsfig}
\usepackage{graphicx}
\usepackage{epstopdf}
\usepackage{amsmath}

\usepackage{rotating}

\usepackage{natbib}
\begin{document}

\title{A Search for Additional Bodies in the GJ 1132 Planetary System from 21 Ground-based Transits and a 100 Hour Spitzer Campaign}
\author{Jason A. Dittmann$^{1}$, Jonathan M. Irwin$^{1}$, David Charbonneau$^{1}$, Zachory K. Berta-Thompson$^{2}$, Elisabeth R. Newton$^{3,}$\footnote{NSF Astronomy and Astrophysics Postdoctoral Fellow}}
\email{Jason.Dittmann@gmail.com}
\affil{[1] Harvard-Smithsonian Center for Astrophysics, 60 Garden St., Cambridge, MA, 02138}
\affil{[2] University of Colorado, 391 UCB, 2000 Colorado Ave, Boulder, CO 80305}
\affil{[3] Massachusetts Institute of Technology, 77 Massachusetts Ave, Cambridge, MA, 02139}

\begin{abstract}
We present the results of a search for additional bodies in the GJ 1132 system through two methods: photometric transits and transit timing variations of the known planet. We collected 21 transit observations of GJ 1132b with the MEarth-South array since 2015. We obtained 100 near-continuous hours of observations with the $Spitzer$ Space Telescope, including two transits of GJ 1132b and spanning 60\% of the orbital phase of the maximum period at which bodies coplanar with GJ 1132b would pass in front of the star. We exclude transits of additional Mars-sized bodies, such as a second planet or a moon, with a confidence of 99.7\%. When we combine the mass estimate of the star (obtained from its parallax and apparent $K_s$ band magnitude) with the stellar density inferred from our high-cadence $Spitzer$ light curve (assuming zero eccentricity), we measure the stellar radius of GJ 1132 to be $0.2105^{+0.0102}_{-0.0085} R_\odot$, and we refine the radius measurement of GJ 1132b to $1.130 \pm 0.056 R_\oplus$. Combined with HARPS RV measurements, we determine the density of GJ 1132b to be $6.2 \pm 2.0$\ g cm$^{-3}$, with the mass determination dominating this uncertainty. We refine the ephemeris of the system and find no evidence for transit timing variations, which would be expected if there was a second planet near an orbital resonance with GJ 1132b.
\end{abstract}

\section{Introduction}
M Dwarfs (0.08 $M_\odot$ $<$ M $<$ 0.60 $M_\odot$) make up approximately 75\% of the stellar population in the Milky Way \citep{RECONS}, and are estimated to host planets similar in size to the Earth ($0.5 R_\oplus < R_p < 1.5 R_\oplus$) at a rate of approximately 1.4 planets per star \citep{Kep_stat_4}. Planets smaller than 1.6 $R_\oplus$ are likely to be primarily composed of rocky terrestrial material \citep{Rogers2015,Dressing_Kepler_93_Mass}. Some planets in this size range appear to have significant gaseous envelopes \citep{Kipping2014}, while some larger planets can be terrestrial in nature \citep{Buchhave2016}, indicating that size may not fully dictate the compositions of planets, and detailed characterization work is required to fully understand their natures. 

The characterization of planets orbiting M dwarfs by the methods of radial velocity and transit techniques is significantly easier than characterizing planets around larger and hotter stars. The RV signal of a planet orbiting a star increases with smaller stellar mass, making rocky planets detectable with measurement precisions demonstrated with current ground-based instrumentation. Similarly, the transit depth (and the size of atmospheric transmission features) increases with decreasing stellar radius. Therefore, characterization of exoplanets is made easiest by studying planets orbiting the coolest M dwarfs.

Due to the cooler temperatures and smaller radii of M dwarfs, the habitable zone is much closer to the star. Habitable planets orbiting M dwarfs have shorter orbital periods, making them easier to discover and observe. An Earth-sized planet around a Sun-like star has a transit probability of only 0.5\%, but this increases to 1.2\% for a planet in the habitable zone of a mid M dwarf. While the $Kepler$ and $K2$ missions have uncovered thousands of extrasolar planets, they operate only on 4.4\% of the total area on the sky (assuming 17 observing fields for K2), and therefore are not surveying the majority of the closest brightest M dwarfs, which are distributed uniformly on the sky. For the closest M dwarf planetary systems, the future Extremely-Large Telescopes (GMT, TMT, and the E-ELT) may be able to search for gases such as O$_2$ at high spectral resolution \citep{Snellen_2013,Rodler_LopezMorales_O2}. JWST has the capability of observing exoplanet transmission spectra at infrared wavelengths, and may be able to identify spectral features of CH$_4$, CO, CO$_2$, and H$_2$O \citep{JWST2,JWST1}.

A new generation of surveys are using robotic array of multiple telescopes to survey the closest mid-to-late M dwarfs for transiting planets. Exoplanet transit surveys like MEarth \citep{Nutzman,jonathan_cool_stars}, TRAPPIST \citep{TRAPPIST_original}, and APACHE \citep{APACHE_start} are currently targeting individual, nearby stars. In the near future, additional surveys like ExTrA \citep{ExTrA} will come online and also target the nearby M dwarfs. The small size and cool temperatures of these stars allow ground based surveys like MEarth to potentially be sensitive to planets in the habitable zone \citep{berta2012}, and for instruments like the $Hubble$ $Space$ $Telescope$ (HST) and the $James$ $Webb$ $Space$ $Telescope$ (JWST) to potentially characterize their atmospheres \citep{gj1214_hubble,jwst_obs_cap}. TRAPPIST \citep{TRAPPIST}, has recently detected three transiting planets orbiting a $0.08 M_\odot$ star 12 parsecs away \citep{trappist1bcd}. This system has already been studied with HST in order to begin obtaining an atmospheric transmission spectrum \citep{trappist_hubble}, and is a high priority target for observations with JWST \citep{trappist_jwst}. 

 The MEarth Project is a photometric survey of mid-to-late M dwarfs estimated to be within 33 pc of the Sun. MEarth aims to detect low mass rocky planets transiting these stars \citep{Nutzman,jonathan_cool_stars}.  MEarth has announced the discovery of two planets, GJ 1214b \citep{gj1214b} and GJ 1132b \citep{GJ1132_discovery}. GJ 1214b has been intensely studied; atmospheric characterization has been performed from optical wavelengths (\citealt{gj1214_rayleigh, gj1214_bean, Murgas}) to the infrared (e.g. \citealt{gj1214_hubble, gj1214_bean, Crossfield_gj1214_spectrum, gj1214_croll, Desert_atmo, Fraine,kreidberg}). Additionally, dedicated, long duration, point-and-stare observations have been performed with the $Spitzer$ Space Telescope in order to search for additional transit planets at periods of 20 days or less \citep{GJ1214_point_and_stare}. These observations did not uncover the existence of another planet in the GJ 1214 system.

Recently, MEarth discovered GJ 1132b, a rocky planet similar in size to the Earth in a 1.6 day orbit around a $0.181 M_\odot$ star 12 parsecs from the Sun \citep{GJ1132_discovery}. At this distance, atmospheric characterization with HST and JWST is feasible. With 4 orbits of HST, low mean molecular weight atmospheres without high-altitude aerosols are detectable. Secondary atmospheres consisting of heavier molecules such as O$_2$ or CO$_2$ require significant investments of observing time ($\approx$60 orbits of HST) to be detected.  GJ 1132b's equilibrium temperature (400 K with a Venus-like albedo of 0.75), makes it unlikely to be habitable, but understanding whether its atmosphere is a primordial or secondary atmosphere can shed light on its formation and migration history and provide key insights into the nature of small planets around M dwarf stars \citep{Schaefer2016}. If additional transiting planets were to be found in this system, then these would also be valuable targets to perform atmospheric transmission spectroscopy measurements, and enable comparative planetology studies to determine under what conditions atmospheres are likely to be primordial or secondary in origin.

Data from the $Kepler$ mission has shed light on the population of exoplanets. \citet{muirhead15} estimated that $21^{+7}_{-5}\%$ of mid M dwarfs host multiple planets, all with orbital periods less than 10 days. Assuming a disk mass to stellar mass ratio of 1\%, these planets would account for nearly all of the non-hydrogen and helium disk mass, and \citet{muirhead15} suggest that this may indicate a very high efficiency in the planet formation mechanism around M dwarfs and a  \emph{lack} of planets at large orbital periods around mid-M dwarfs. A study by \citet{Ballard16} generated synthetic populations of exoplanets described by a range of planet multiplicity and mutual inclination for comparison to the $Kepler$ M dwarf sample. They found that the total number of single transiting planet systems is incompatible with the number of systems with multiple transiting planets assuming a single planet population distribution. Instead, half of all M dwarf planetary systems consist of a single planet, while half of M dwarf systems contain 5 or more planets with a low mutual inclination \citep{Ballard16}. 

A theoretical understanding of the origin of this Kepler dichotomy has focused on multiple physical variables within the protoplanetary disk. \citet{Moriarty_and_Ballard} found evidence that the distribution of short period planets are frozen-in during the formation process and are long-term stable and are not a result of metastable systems becoming unstable later in their lifetimes. The $Kepler$ dichotomy then originates from the surface density profile of the protoplanetary disk in which these planets form. The surface density profile of protoplanetary disks may also determine whether a system forms planets that are primarily rocky or whether planets with significant gaseous envelopes are formed \citep{Dawson_dichotomy}. Disks with higher surface density of solids are able to form large rocky bodies while the gas disk still exists, enabling these bodies to accrete and maintain primordial gaseous atmospheres. Disks with a smaller surface density of solids take longer to form solid rocky bodies, and gas disk is more likely to dissipate before these bodies become large enough to accrete gas. \citet{Dawson_dichotomy} conclude that M dwarfs with super-solar metallicities are more likely to host planets with gaseous atmospheres while those with sub-solar metallicities are more likely to form tightly packed, small, rocky planets without primordial gaseous atmospheres.

Here we describe observations of the GJ 1132 system taken with the MEarth Observatory and the $Spitzer$ Space Telescope. We have collected 21 transit observations with MEarth in our red optical bandpass and have observed for 100 near-continuous hours with $Spitzer$, collecting two transits. We seek to probe the GJ 1132 system for signs of additional transiting bodies in the system, including moons orbiting GJ 1132b and additional planets. In section 2, we describe the observations from MEarth-South and from $Spitzer$. In section 3, we describe our data analysis techniques and modeling the transit light curves of GJ 1132b. In section 4, we discuss the photometric precision of our $Spitzer$ data set and our sensitivity to additional transiting bodies. 

\section{Observations and Production of the Time Series}

\subsection{MEarth-South}
MEarth-South consists of eight f/9 40 cm Ritchey-Chr\'etien telescopes on German equatorial mounts situated at Cerro Tololo International Observatory (CTIO) in Chile \citep{jonathan_cool_stars}. The telescopes are robotic and take data on every clear night. Each telescope has a 2048 $\times$ 2048 pixel CCD, with a pixel scale of approximately 0.84\arcsec pix$^{-1}$. We use a Schott RG715 glass filter with an anti-reflection coating, which has a broad band red optical throughput. Each CCD is an e2v CCD230-42 device with an interference coating matched to out bandpass, which has the effect of reducing fringing. The CCDs operate at $-30^{\circ}$C. Prior to each exposure we pre-flash the detector to eliminate persistence from the previous exposure. This increases the dark current in the CCD, which we subsequently subtract off in image processing.

We gather sky flat fames at dawn and at dusk each night.  Since MEarth uses German equatorial mounts, we must rotate the field-of-view of the detector by 180 degrees relative to the sky when crossing the meridian. Therefore, we take two sets of flat fields each at dawn and at dusk, taking adjacent pairs of flat fields on opposite sides of the meridian. These flats allow us to average out large-scale illumination gradients from the Sun and the Moon. Scattered light in our telescopes concentrates in the center of the field of view with an amplitude of approximately 5\% of the average value of the sky across the CCD. To correct for the scattered light, we filter out large scale structure in the background from our combined twilight flat field and use the residual flat field to track small scale features (inter-pixel sensitivity and dust shadows). The large scale flat field is derived from dithered photometry of dense star fields.

We measure the nonlinearity of the MEarth detectors with a dedicated sequence of dome flats. The nonlinearity of the detectors is approximately 1\%-2\% at half the detector full well and rises to 3\%-4\% near saturation. We correct for this nonlinearity as part of the general MEarth pipeline and all exposure times are set to avoid surpassing 50\% of the full well.

In order to reach the precisions necessary to detect exoplanetary transits, we must measure and correct for the effects of differential color extinction due to the atmosphere. Changes in the amount of precipitable water vapor in the atmosphere over the course of an observing night results in changes in the atmospheric transmission in the red end of the MEarth bandpass. Since the MEarth targets are, by their nature as mid-to-late M dwarfs, the reddest objects in the field, these effects can produce systematic effects when using bluer stars as reference stars, since our target stars are more sensitive to the changes in the precipitable water vapor than the field reference stars. We correct for these effects by measuring a ``common mode" for all of our target M dwarfs. The common mode is the measure of the average differential light curve of all M dwarfs currently being observed by MEarth-South, in time bins of 0.02 days (28.8 minutes). This correlated behavior between targets is a good proxy of the local changes in precipitable water vapor through the night. In order to measure this common mode, half of the MEarth-South telescopes must be observing other M dwarfs during dedicated transit observations. Therefore, for each observed transit presented here, either 3 or 4 of the MEarth-South telescopes are observing GJ 1132; the rest are observing other M dwarfs to determine the common mode. We targeted all transits of GJ 1132b visible from CTIO between 05 November 2015 and 05 July 2016 civil date. We observed a total of 21 transits with 6 different telescopes over this time period and obtained 34968 data points (we also observed some partial transits, but they are not included in this analysis).

We measure the position of the target and reference stars on each frame using a modified method from \citet{I85}. We bin each image into 64 $\times$ 64 pixel blocks and measure the peak of the histogram of the intensity of these super-pixels as an estimate of the local sky background. This process eliminates large scale illumination gradients in image background. We estimate the sky background around each star with a sky annulus between 18 and 24 pixels away from the stellar photo center. The photo center of each star is determined from the intensity weighted first moment (the centroid) of the star.

We measure the total flux using a 6 pixel ($\approx 5.04\arcsec$) or 8 pixel ($\approx 6.72\arcsec$) aperture radius, depending on the seeing conditions of the individual night and adopt an aperture correction to correct for the stellar flux that lies outside of the aperture. The aperture correction varies from night to night with atmospheric conditions, but has a typical value around 0.04 magnitudes. As we are performing relative photometry for this analysis, the aperture corrections of the target star and the reference stars cancel out. Pixels that are partially within the circular aperture are weighted by the fraction of the pixel that would lie within an idealized circular aperture at the stellar location. 

Each transit observation from each telescope is reduced independently of the others, although the common mode (see observation section above) is common to all telescopes. Each light curve out-of-transit baseline is normalized to 1.0 and we fit a linear term times the common mode to the time series to remove trends due to the atmosphere in the data. We vary each eclipse normalization in our light curve analysis but we do not vary GJ 1132's coupling constant to the common mode during our analysis.

\subsection{$Spitzer$}
$Spitzer$ obtained data with the Infrared Array Camera (IRAC) at 4.5 $\mu$m as program 12082 (PI Dittmann). We obtained 100 hours of nearly-continuous $Spitzer$ observations, beginning on 24 April 2016 and ending on 28 April 2016 UT. Observations spanned from BJD 2457502.0064004 to 2457506.3296307. During this span, there were 7 breaks in the data longer than 10 seconds in duration. The largest, between BJD 2457503.4590452 and 2457503.6895174 (approximately 5.5 hours), was due to a data downlink.

Our program was divided into 6 Astronomical Observation Requests (AORs). The duration of these AORs were 20 hours, 14 hours, 20 hours, 20 hours, 15 hours, and 8 hours. Between each AOR there was a gap of 12 to 250 seconds with the 5.5 hour downlink occurring at the end of the 14 hour AOR. This data set consists of 2725 sets of 64 individual sub array images with an integration time of 2 seconds, for a total of 174400 total data points. We obtained data with the subarray mode, placing GJ 1132 in the portion of the detector that is well characterized for the purpose of obtaining high precision light curves. In order to improve cadence and have a reasonable data volume, we utilized a small 16 $\times$ 16 pixel portion of the detector. These data were calibrated with the $Spitzer$ pipeline version S19.2.0, and the timestamps of each data point are calculated at the Solar System barycenter.

We correct our Spitzer data with the pixel-level decorrelation (PLD) method, described by \citet{Deming_PLD}. The approach of the PLD reduction method is that it uses the brightness values encoded on the pixels themselves instead of correlating brightness fluctuations with a measure of the location of the star on the CCD (which itself depends on the pixel values). Due to the pointing stability of $Spitzer$, the location of the stellar image does not drift significantly, even over observations lasting several hours. The 50th percentile of the photocenter difference from the median photocenter in each AOR is 0.062 pixels. The 95th percentile for the difference in the photocenter from its median location is 0.126 pixels. At the beginning of each AOR, the target falls on a slightly different location of the array, but the photocenter remains stable within each AOR. This pointing stability and the relatively large size of the $Spitzer$ point spread function (PSF) allow us to describe the instrument-based variations in the lightcurve from image motion and pixel sensitivity as a linear combination of the pixels in the image themselves.

We select a $5 \times 5$ pixel area centered on the center pixel of the subarray encompassing the flux from GJ 1132. We sum the pixels in this square area in order to obtain the total brightness in this aperture and then divide each pixel by this value in order to normalize each pixel to this value. We do this for each of the 174400 images. \citet{Deming_PLD} note that binning the data prior to fitting pixel coefficients can provide better stability on timescales relevant for planetary transits in exchange for poorer stability on shorter (several second) timescales. Binning data allows us to better determine the pixel coefficients at the edge of the aperture, where the flux is low in any individual 2 second exposure. We bin our data into 60 second blocks for the purposes of fitting our model coefficients. We note that since this method involves normalizing the pixel values as a percentage of the total flux, it removes astrophysical variations (i.e. the transit) from this normalization procedure, while allowing variations due to pixel sensitivity, flat fielding error, and image motion to be calibrated out via the relative values of each pixel coefficient.

We fit the following model:
\begin{equation}
F_i =  \sum_{n=1}^{25} C_n P_{n,i} + b
\end{equation}

where $F_i$ is the total flux at time $i$, $n$ is the pixel number (of the 25 pixels in the model), $C_n$ is the coefficient for each pixel, $P_{n,i}$ is the normalized value of pixel $n$ at time $i$, and b is a constant. Since the pixel values are normalized and for the purpose of this model, the variations in $F_i$ are assumed to be due to the flux from individual pixels shifting to adjacent pixels as the photocenter shifts during the observation. 

 \citet{Deming_PLD} also included a linear and a quadratic term in time in their model, but we omit those terms here as we find them to be insignificant for our dataset. When fitting this model, we eliminate any data points where the total unnormalized flux is more than $3 \sigma$ discrepant from the median across the AOR. Since we are seeking to obtain only the coefficients for each pixel, this will mitigate the potential influence of individual outliers. We also exclude all in-transit data points from this analysis, although a reduction including these data points does not significantly affect our results and therefore we do not think that this data reduction method can suppress potential transit signals from other bodies in the system. This is because the pixels are normalized for each time stamp, so real astrophysical variations are eliminated. The only way for this reduction method to suppress real astrophysical variations is if they are directly correlated with shifts of the stellar photocenter. We fit each of our 6 individual AORs independently. The pixel coefficients can change by as much as 50\% due to the new average location of the photocenter with each repointing. 

Once we obtain the pixel coefficients for each individual AOR, we apply these coefficients to the \textit{unnormalized} pixel level data, and sum the data in these pixels in order to obtain the flux of GJ 1132. All data that was omitted for the purpose of fitting the pixel coefficients is reinstated for the purpose of measuring the light curve. We apply a new outlier rejection method to these data. For each individual data point, we measure the median flux value within a $\pm$10 minute window of that data point as well as the median absolute deviation from the median (MAD, \citealt{Stat_book}) in this window. If the data point is more than $10$ MADs discrepant from the local median value, it is discarded; 169 of 174400 data points are discarded due to this criterion. None of these excluded data points occur within 10 minutes of each other, and therefore we believe these to be systematic outliers and not indicative of a short timescale astrophysical variability.  

In Table \ref{Photometry_table}, we provide the corrected photometry for GJ 1132 for all eclipses taken with MEarth-South and $Spitzer$.

\section{Modeling of the Time Series}

We  obtained observations of 21 transits in the near red optical MEarth filter and 2 transits in the $Spitzer$ 4.5 $\mu$m channel. Each MEarth transit contains both pre-transit and post-transit data for the purpose of establishing an out of transit baseline, while the $Spitzer$ dataset includes approximately 98 hours of out-of-transit observations. We fit our observations with the \texttt{batman} code \citep{batman}, which is an an optimized, python implementation of the \citet{mandel_and_agol} analytic model for transit light curves. In Table \ref{model_parameters} we describe the parameters of the model we use to fit our transit observations.

We initiate our model with the physical parameters found in \citet{GJ1132_discovery}. We adopt limb darkening coefficients from \citet{claret2012_limb_darkening} for a 3300 K star with a log($g$) = 5.0. We adopt the limb darkening coefficients for a Cousins $I$ filter for our MEarth observations, as the effective wavelength is similar. For our $Spitzer$ data, we adopt limb darkening parameters from \citet{Spitzer_Claret}, which are calculated for the $Spitzer$ bandpass. Due to the large quantity of data compared to the initial discovery observations, we adopt very loose priors for our model. For our priors, we let each parameter vary freely with no penalty within $5\sigma$ of the values determined by \citet{GJ1132_discovery}. We let limb darkening coefficients vary freely (uniform prior) within 20\% of their initial value from \citet{claret2012_limb_darkening}, to account for slight differences between GJ 1132 and the stellar models as well as for differences in the effective bandpass between the $MEarth$ and the Cousins $I$ filter. We fix the eccentricity of GJ 1132b to zero. While fixing the eccentricity to zero may bias the stellar density measured from the light curve \citep{carter2011gj1214}. However, the short period of GJ 1132b implies that the eccentricity is either 0 or very low due to tidal forces (circularization timescale of approximately 400,000 years \citealt{GJ1132_discovery}), and would have little effect on our measurement. The radial velocity measurements obtained by \citet{GJ1132_discovery} were unable to provide a robust constraint on eccentricity. Our choice of priors does not significantly affect our result, as the parameters in our chain to not vary to this extent. 

In order to explore parameter space, we use the \texttt{emcee} code \citep{emcee}, a python implementation of the Affine Invariant Markov Chain Monte Carlo sampler. We allow the ratio of the radii to vary between the different bandpasses. This allows us to probe the effect of unocculted starspots (which will more strongly affect the red-optical MEarth data than the longer wavelength $Spitzer$ data), as well as begin to probe the transmission spectrum of GJ 1132b's atmosphere. 

\begin{figure}
\begin{singlespace}
\centering
\includegraphics[width=1.0\linewidth]{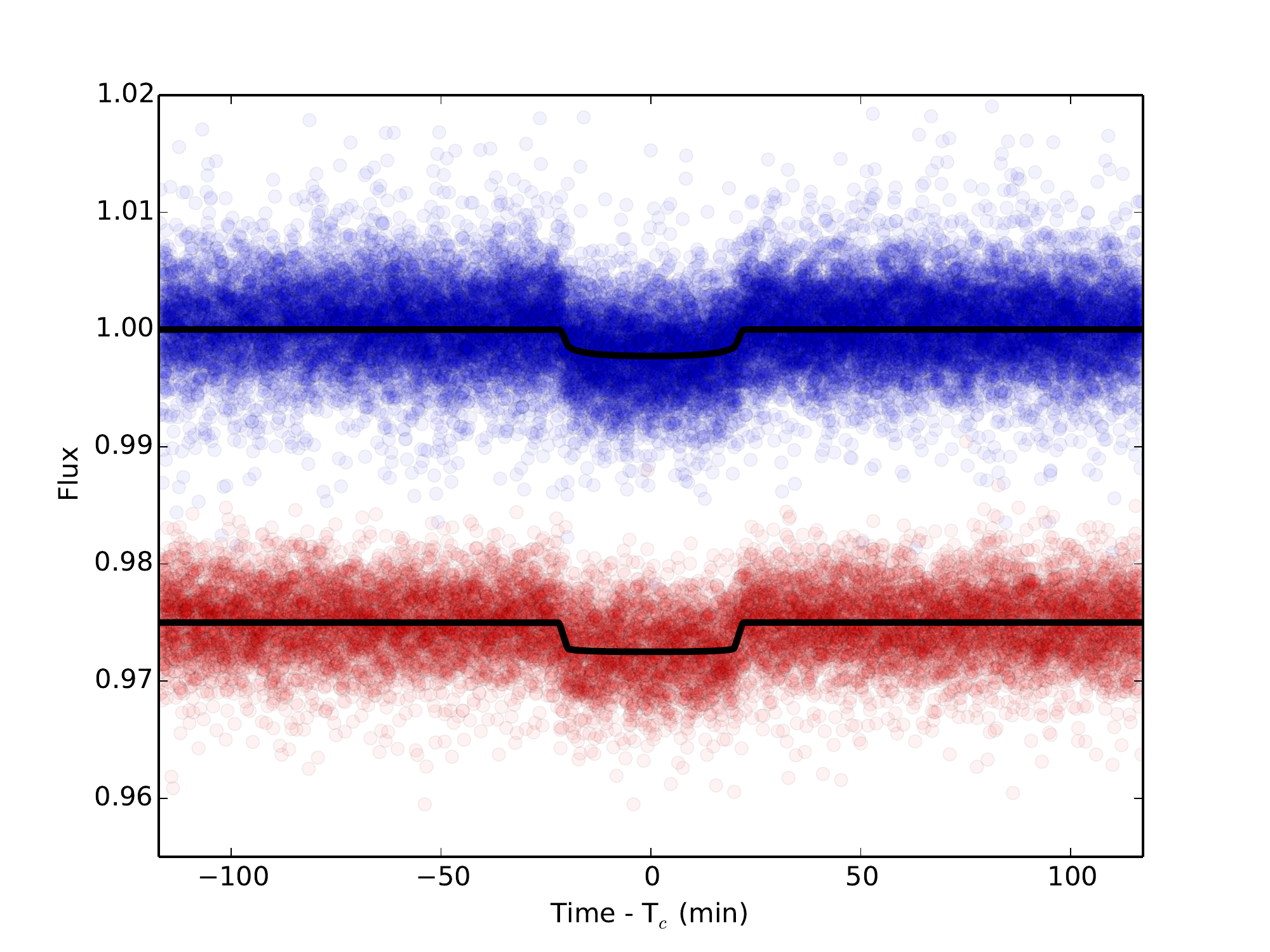}
\caption{Stacked transit light curves of 21 transits from the MEarth Observatory (blue) and 2 transits from $Spitzer$ (red, offset for clarity). The black line is our best model fit to the data. We find a slightly larger planet to star radius ratio in the MEarth red-optical passband than in the 4.5$\mu$m $Spitzer$ passband, which we attribute to the effect of unocculted star spots. We use these transit light curves to refine the orbital ephemeris of the planet, as well as resolve the ingress and egress times, placing stronger constraints on the radius of the star (and therefore the radius of the planet), than previously determined.
}
\label{transit_curves}
\end{singlespace}
\end{figure}

\begin{figure}
\begin{singlespace}
\centering
\includegraphics[width=1.0\linewidth]{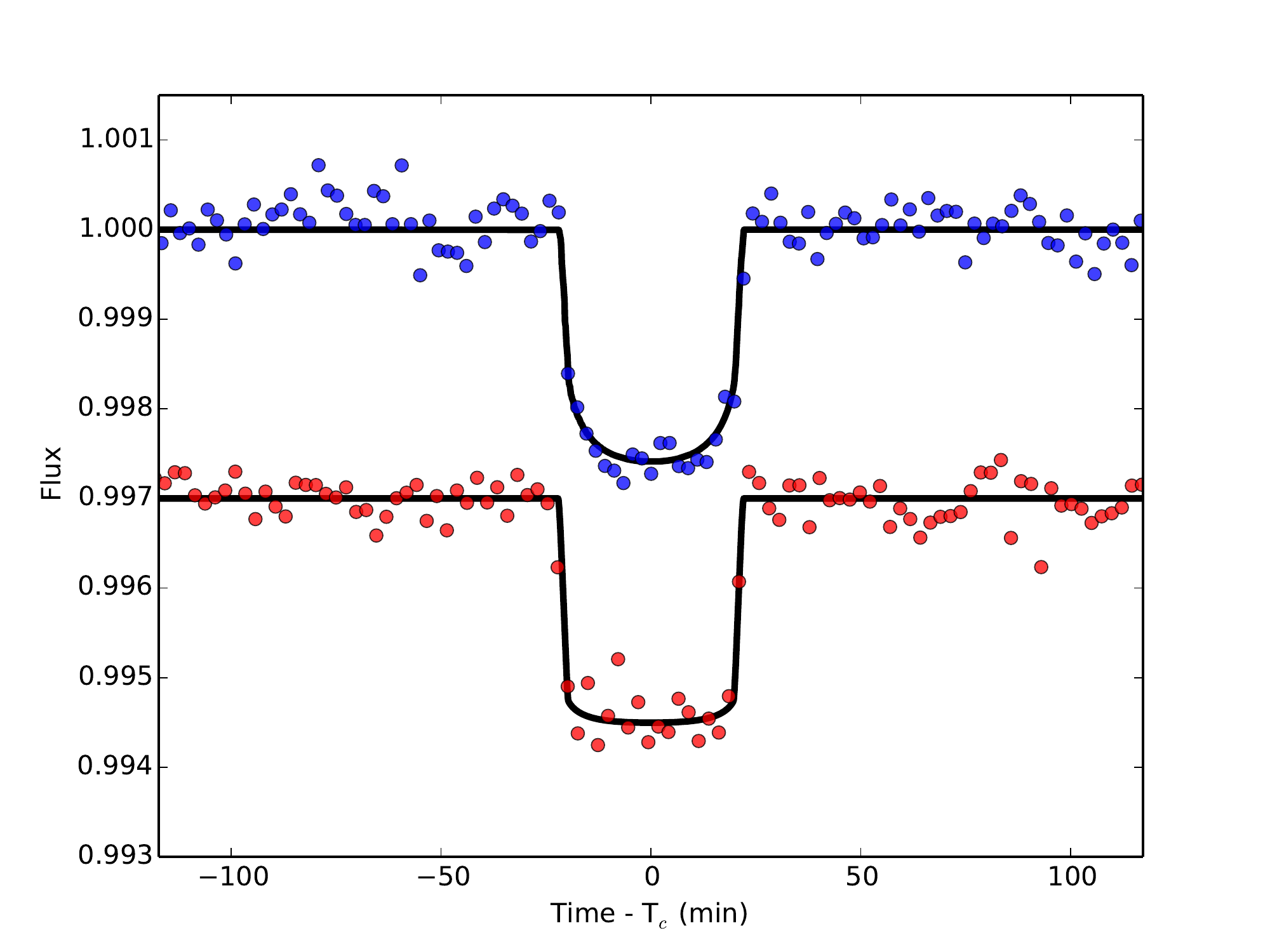}
\caption{MEarth (blue) and $Spitzer$ (red) observations of transits of GJ 1132b, binned on 3 minute time scales (90 data points per bin, for $Spitzer$), offset for clarity. We measure a transit depth of $2.42 \pm 0.08$ mmag. The high-cadence capabilities of $Spitzer$ allow us to resolve the ingress and egress time, allowing us to more reliably measure the parameters of the system. }
\label{Spitzer_zoom}
\end{singlespace}
\end{figure}

Each model is initiated with 100 walkers in a Gaussian ball located at the initial solution described above. We run each chain for 75000 steps and discard the first 10\% of the resultant samples so that the solution may ``burn-in" irrespective of the initialization. This burn-in appears to occur earlier than this cut-off for all parameters in the model, but we retain this 10\% cut-off to avoid any possible systematics. We report the best fit model from this chain, as well as the 16th and 84th percentile for each parameter in Table \ref{model_fit}. In Figure \ref{transit_curves}, we show the transit data collected by MEarth and $Spitzer$ as well as our best fitting model, and in Figure \ref{Spitzer_zoom} we show solely the transits observed by $Spitzer$, binned on 3 minute time scales, as these observations are more precise and have sufficient cadence to resolve the ingress and egress of the transit. 

Our best fit measurement for the ratio of the planetary and stellar radii is $\frac{R_p}{R_*} = 0.0455 \pm 0.0006$ in the MEarth bandpass and $\frac{R_p}{R_*} = 0.0492 \pm 0.0008$ in the $Spitzer$ $4.5\mu$m channel. The MEarth-bandpass, being bluer, is more sensitive to the effects of starspots on the star. While GJ 1132 is a photospherically quiet star, some magnetic activity and star spots exist on its surface, as we have been able to measure rotational modulation due to the longitudinally asymmetric distribution of these spots \citep{GJ1132_discovery}. In order to reconcile the difference between the observed transit depth in $MEarth$ and $Spitzer$, GJ 1132b's transit chord must lie along an active stellar latitude. For a starspot that is completely dark in the MEarth bandpass compared to the non spotted stellar surface, GJ 1132b must occult a starspot only 540 km in radius in order to account for the observed transit depth difference. For a starspot with an effective temperature 0.7 times the effective temperature of the non-spotted photosphere, the starspot size required increases to 1000 km. This is consistent with the starspot size distribution of NGC 2516, although there are significant degeneracies between starspot size, starspot filling fraction, and the starspot to photosphere temperature ratio \citep{Jackson2013_starspot_size}. We believe that the effects of planetary starspot occultation can explain the difference between these two radii measurements and are not due to a broad spectral feature in GJ 1132b's atmospheres, which would require a scale height 540 km deeper in the blue MEarth bandpass than at the $Spitzer$ 4.5 $\mu$m bandpass. 

\begin{figure}
\begin{singlespace}
\centering
\includegraphics[width=1.0\linewidth]{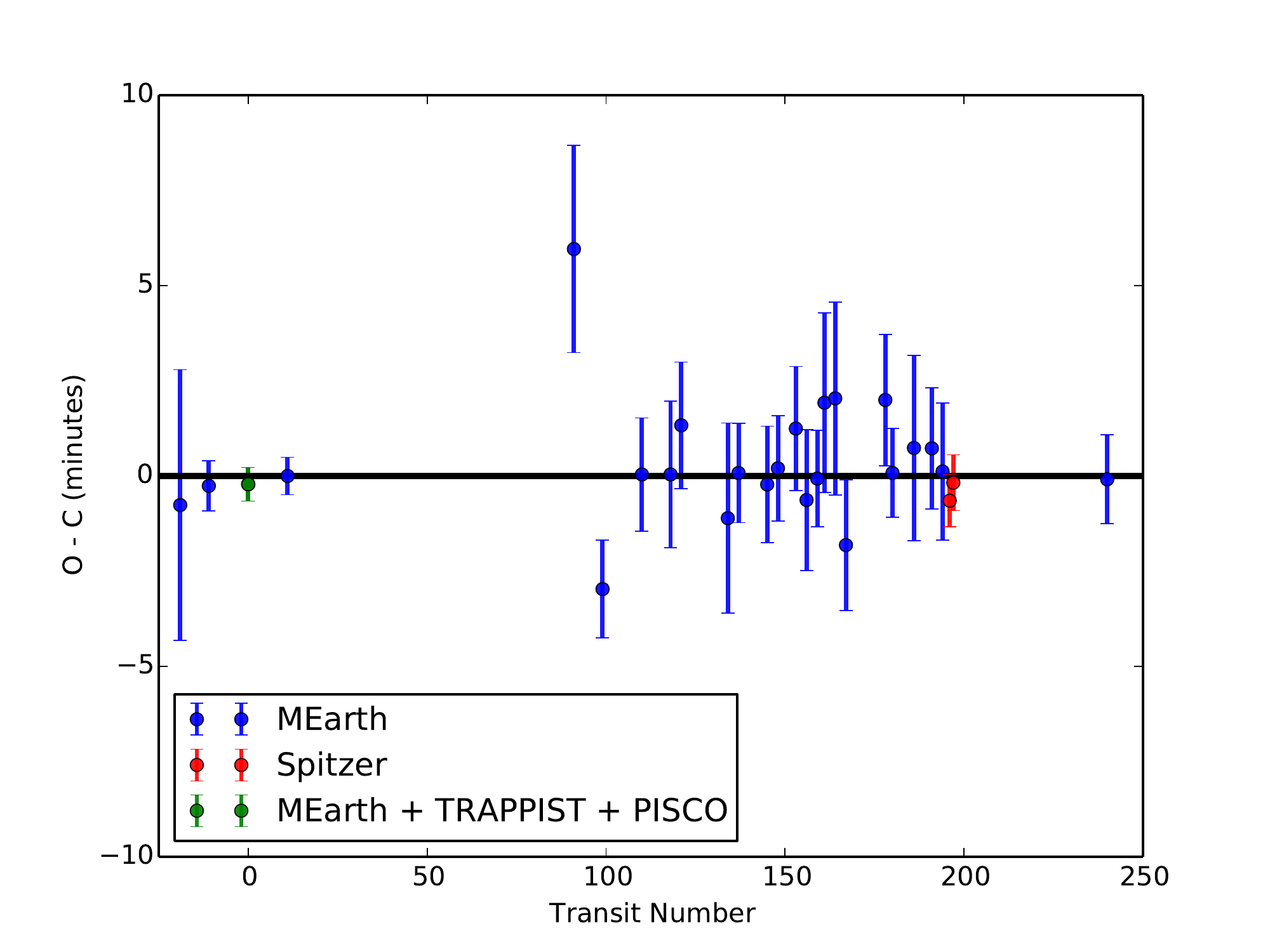}
\caption{Observed - Calculated (OC) diagram for the transits presented in this paper. Blue data points represent measurements from MEarth-South, the green data point at epoch 0 is the measurement from a combination of MEarth-South, TRAPPIST, and PISCO data presented by \citet{GJ1132_discovery}, and the red data points represent the measurements from $Spitzer$. We measure the best fitting individual transit time for each transit using our globally best-fitted transit model, varying only the central time of transit. We measure the difference between the transit time of each individual transit and the linear ephemeris from our best fit model. We find no evidence of transit timing variations in this system, suggesting that any additional bodies in the GJ 1132 system are not in or near mean motion resonances with GJ 1132b. 
}
\label{OC_diagram}
\end{singlespace}
\end{figure}

Additionally, with our best fit transit model we fit each transit individually solely for the best transit time, in order to potentially detect transit timing variations due to the presence of a perturbing body. We measure each individual transit time by holding our best fit transit model constant, and solely fitting the central time of transit. We construct an observed - calculated diagram (OC) showing the deviation of an individual transit time from the best fit ephemeris, including the transits measured in the initial discovery observations in \citet{GJ1132_discovery} (Figure \ref{OC_diagram}), and provide our individual measured transit times in Table \ref{timeing_table}.

\section{Discussion and Conclusion}

\subsection{Physical Parameters of the GJ 1132 system}
With a high precision, high cadence light curve, we are able to refine the measured parameters of this system. In particular, the 2-second cadence of the $Spitzer$ frame allows us to resolve the ingress and egress duration of the transit, which was not possible with the ground based discovery data. This allows us to directly measure the stellar density. We use the stellar mass determined by \citet{GJ1132_discovery}, which relied on a trigonometric parallax from \citet{SN13} and the mass-luminosity relation of \citet{Delfosse}, combined with our measurement of the stellar density, to measure the radius of GJ 1132. We measure a stellar radius of $R = 0.2105^{+0.0102}_{-0.0085}$ $R_\odot$, which is consistent with the value originally reported by \citet{GJ1132_discovery}. We reiterate that this value may be biased by our assumption of zero eccentricity for the orbit of GJ 1132b \citep{carter2011gj1214}, but believe that a 0 or negligible eccentricity is likely due to the close-in orbit of GJ 1132b. We use this measurement of the stellar radius as well as our measurement of the transit depth from the $Spitzer$ light curves, which are less affected by the effects of starspots, to measure a planetary radius of $1.130 \pm 0.056 R_\oplus$, consistent but more precise than the value determined by \citet{GJ1132_discovery}.

\begin{figure}
\begin{singlespace}
\centering
\includegraphics[width=1.0\linewidth]{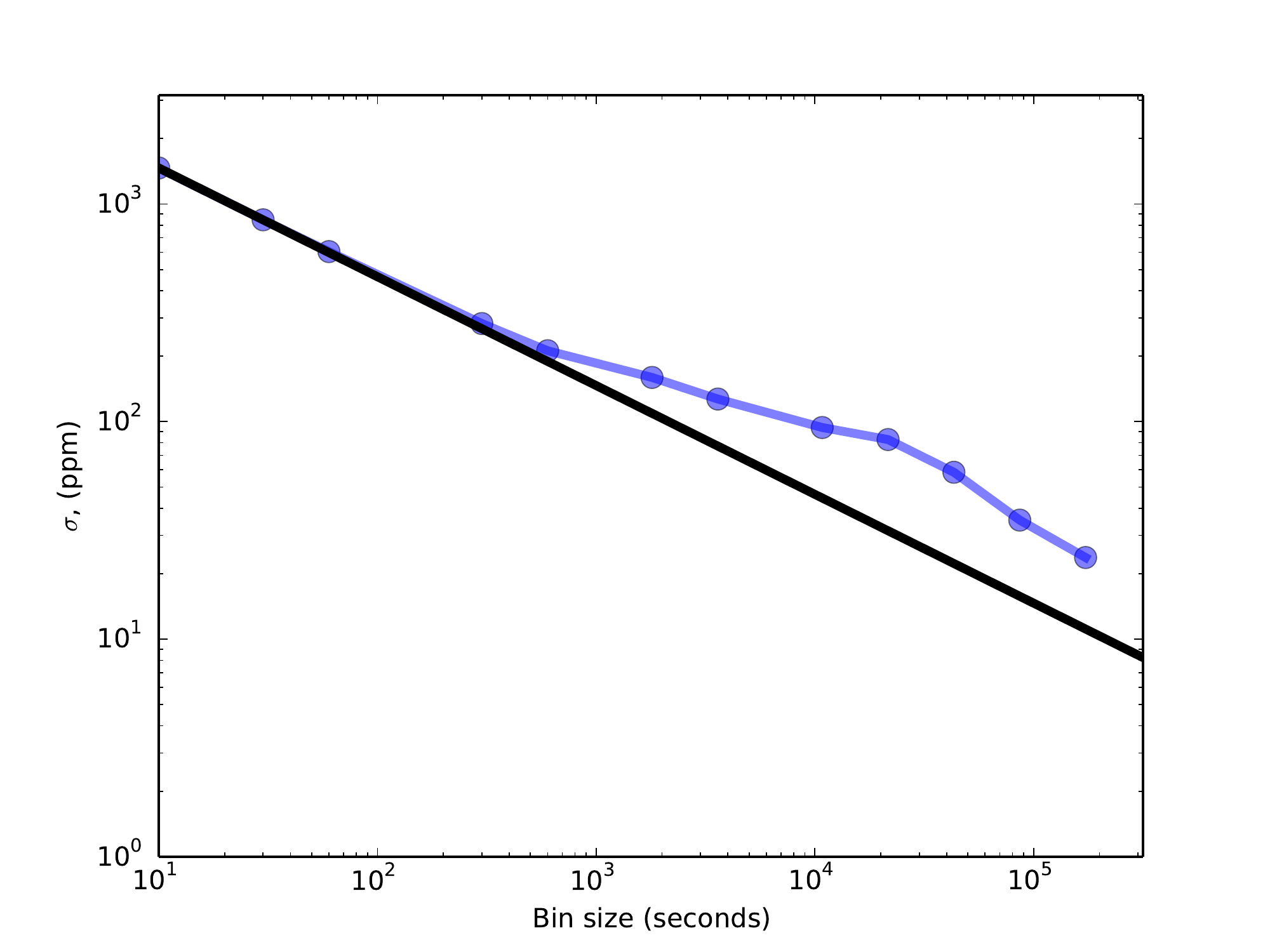}
\caption{Standard deviation of the $Spitzer$ 100 hour time series as a function of bin size. The black line is a line of slope -1/2 anchored to measured standard deviation in 10 second bins. We find that our $Spitzer$ data and reduction technique follows the photon noise limit to a precision of 250 ppm, which we achieve at 10 minute bin sizes. We reach a photometric precision of 159 ppm in 30 min, which is slightly shorter than the duration of the transit of GJ 1132b and similar to the transit duration that would be expected for planets interior to GJ 1132b. For a 0.21 $R_\odot$ star and a $3\sigma$ detection of the transit depth, this corresponds to a body 94\% the radius of Mars.
}
\label{photometric_stability}
\end{singlespace}
\end{figure}

\citet{Deming_PLD} demonstrated using $Spitzer$ data taken of the M dwarf planetary host GJ 436 \citep{Ballard_Spitzer} that they could recover near to photon limited behavior down to a level of 100 ppm in 1000 seconds. We find similar behavior for the PLD algorithm for our data set. In Figure \ref{photometric_stability} we show the standard deviation of our residuals from the $Spitzer$ observations as a function of bin size. We find that our $Spitzer$ data shows approximately photon limited behavior for bin sizes of 10 minutes and smaller, for a precision of approximately 200 ppm. At larger bin sizes we do not recover photon limited behavior, although the standard deviation of our residuals continues to decrease. On 30 minutes timescales we obtain a standard deviation of 160 ppm, for a $1\sigma$ transit precision equal to the size of Earth's moon. 

\subsection{Limits on Single Transit Events from Other Bodies}
The observations presented here contain 100 hours of observations in a 105 hour window, with the only significant gap in observations occurring during an Earth data downlink. 
We find an orbital inclination of $88.68^{+0.40}_{-0.33}$ degrees for GJ 1132b. Assuming coplanarity, additional planets in the system would also transit GJ 1132 out to a period of 6.9 days, longer than the observations presented here. Therefore, if any coplanar planetary bodies exist in the GJ 1132 system with periods of 4 days or less, we should see them transit during our set of observations. Between 4.17 days (100 hours) and 6.9 days, our sensitivity decreases due the increasing probability that a potential transit falls outside the $Spitzer$ window of observations. At periods longer than 6.9 days, coplanar objects can still transit, but this requires a nonzero eccentricity and alignment of the line of nodes.

\begin{figure}
\begin{singlespace}
\centering
\includegraphics[width=1.0\linewidth]{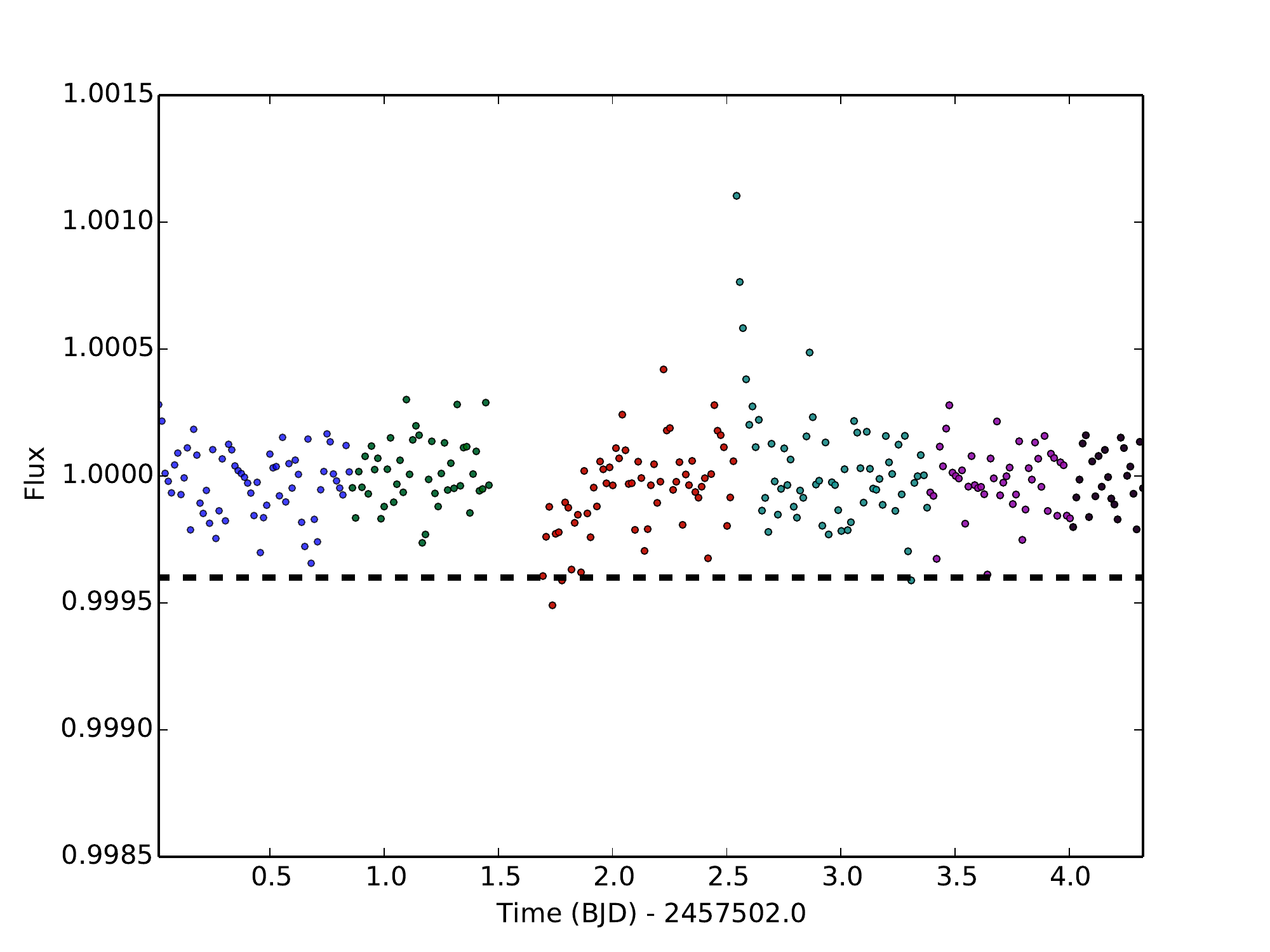}
\caption{Flux of GJ 1132 with transits of GJ 1132b removed (residuals are shown) for the 100 hour $Spitzer$ campaign. Each color represents a new AOR, where a $Spitzer$ repointing occurred. We have binned the data on 20 minute time scales. We reach a photometric precision of 159 ppm in 30 minutes (35\% shorter than the transits of GJ 1132b). With this precision we would detect the transits of a body the size of Mars with $3 \sigma$ confidence (shown as black dashed line). We note that a transit would occur over multiple consecutive bins, and that while individual bins cross this threshold, we do not detect a transit-like event. We find no evidence for additional transiting bodies during these observations and detect zero $3\sigma$ negative outliers during the these observations. The negative outliers at the start of the AOR after the Earth-pointing data downlink are likely due to thermal systematics from the data downlink and the large change in spacecraft orientation and are unlikely to be astrophysical in origin (see section 4.5). The brightening event visible at BJD 2457504.5 occurs at the beginning of an AOR, and we think it is likely to be a systematic effect and not due to a flare from the host star. 
}
\label{spitzer_lightcurve}
\end{singlespace}
\end{figure}

In Figure \ref{spitzer_lightcurve}, we show the flux of GJ 1132 at $4.5 \mu$m during the entirety of the $Spitzer$ campaign, binned on 20 minute timescales (roughly half the duration of a transit), with transits of GJ 1132b subtracted with our best fitting transit model (residuals from this fit are shown). While there are no obvious transits of large bodies visible in this data set, we attempt to assess our sensitivity to single transits using the observed standard deviation of our data set on timescales relevant to exoplanetary transits. In Figure \ref{photometric_stability} we show that we can recover photon limited behavior to a time scale of 10 minutes, at $200$ ppm precision. In order to identify possible in-transit events, we bin our data to 10-minute time scales and search for any negative outlier detected at 3$\sigma$, corresponding to a sensitivity to bodies the size of Mars. We have 590 data bins. With white noise fluctuations alone, we would expect 1.8 $3 \sigma$ outliers from white noise fluctuations alone, half of which would be negative outliers. Since a transit signal is likely to extend for greater than 10 minutes in duration, we would expect any significant transit signal to span more than one consecutive bin. However, we find 0 $3 \sigma$ negative outliers in this data set.

\begin{figure}
\begin{singlespace}
\centering
\includegraphics[width=1.0\linewidth]{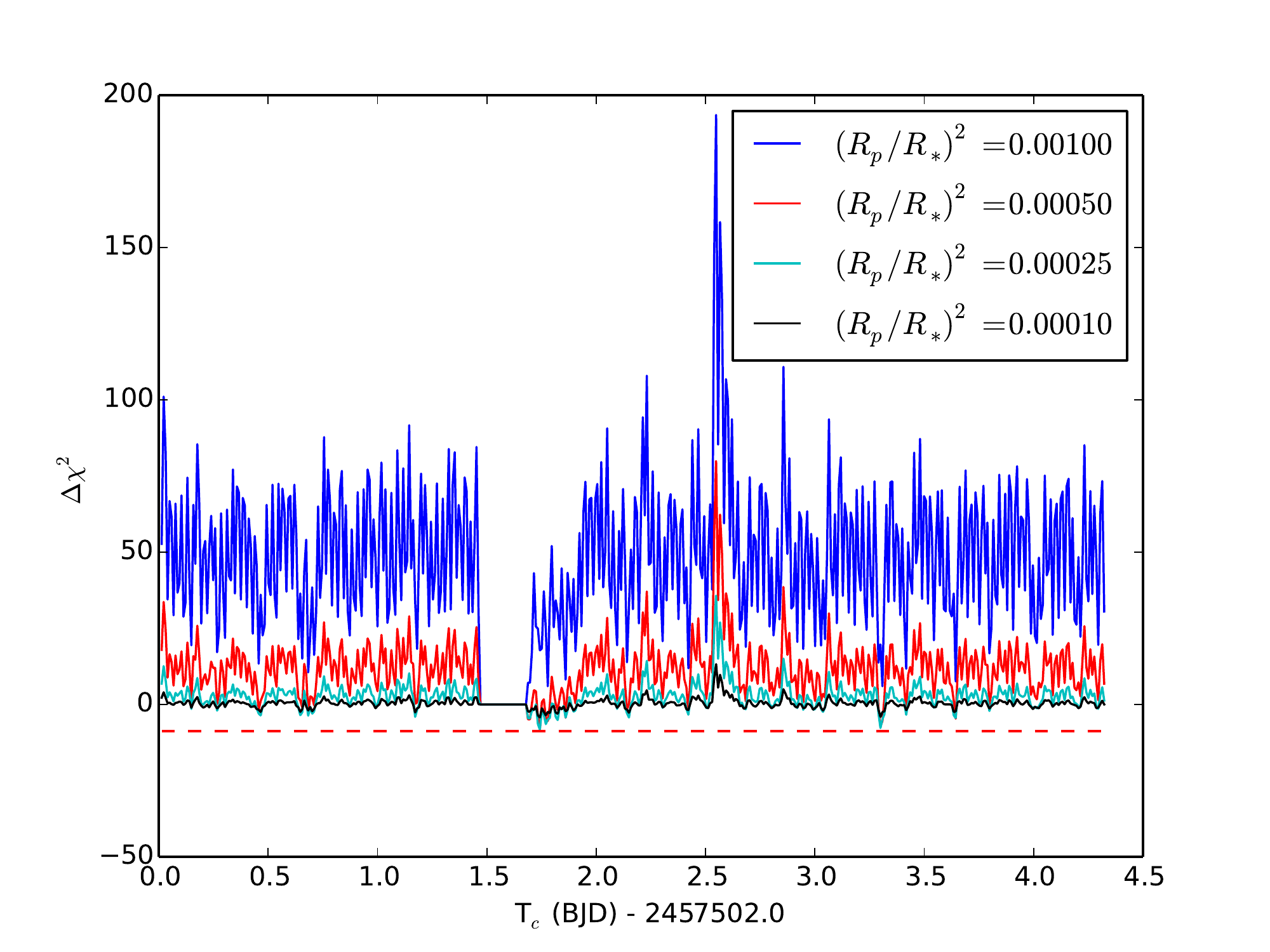}
\caption{$\Delta \chi^2$ vs central transit time for box-model transits compared to GJ 1132b-only models. Negative $\Delta \chi^2$ indicate a better fit to the data. The lower values after the data downlink are due to systematic effects associated with the data downlink, while the large spike around 2457504.5 BJD is due to a systematic associated with a repointing events. We find no significant box-like transit signals in this dataset, and can exclude signals belonging to a body 0.85 $\times$ the size of Mars and larger. The approximate $\Delta \chi^2$ we would expect from the transit of a body 0.85 $\times$ the size of Mars is indicated by the dashed red line. 
}
\label{delta_chi_2_plot}
\end{singlespace}
\end{figure}

We estimate the minimum radius transiting body we can exclude with this dataset by comparing the $\chi^2$ of our transit model with solely GJ 1132b and a model that includes a box-model transit of another body. We choose a box width of $40$ min, which is a similar timescale to that we would expect from a transiting body. We vary the depth of the box in order to assess our sensitivity to other bodies. In Figure \ref{delta_chi_2_plot}, we plot the $\Delta \chi^2$ from box models of varying transit depth and choice of central transit time when compared to our GJ 1132b-only model. Better fits to the data have a negative $\Delta \chi^2$. We find that that we can exclude transiting bodies $0.85$ times the size of Mars or larger with orbital period of 100 hours or less. This observation also excludes 60\% of the orbit of bodies located at 6.9 day orbital periods (the maximum period a coplanar object with zero eccentricity would still transit the host star). Smaller bodies are permitted, as they do not significantly change the $\chi^2$, but we see no evidence for them in this data set. Since this data also spans the times around the transits of GJ 1132b, we can also exclude exomoons around GJ 1132b to this same size. However, we note that due to the proximity of GJ 1132b to its host star, the hill sphere of GJ 1132b overlaps with the Roche lobe. Therefore, exomoons are not dynamically stable around GJ 1132b and we do not expect any to exist.  

\subsection{Limits on Extremely Short Period Bodies}
\begin{figure}
\begin{singlespace}
\centering
\includegraphics[width=1.0\linewidth]{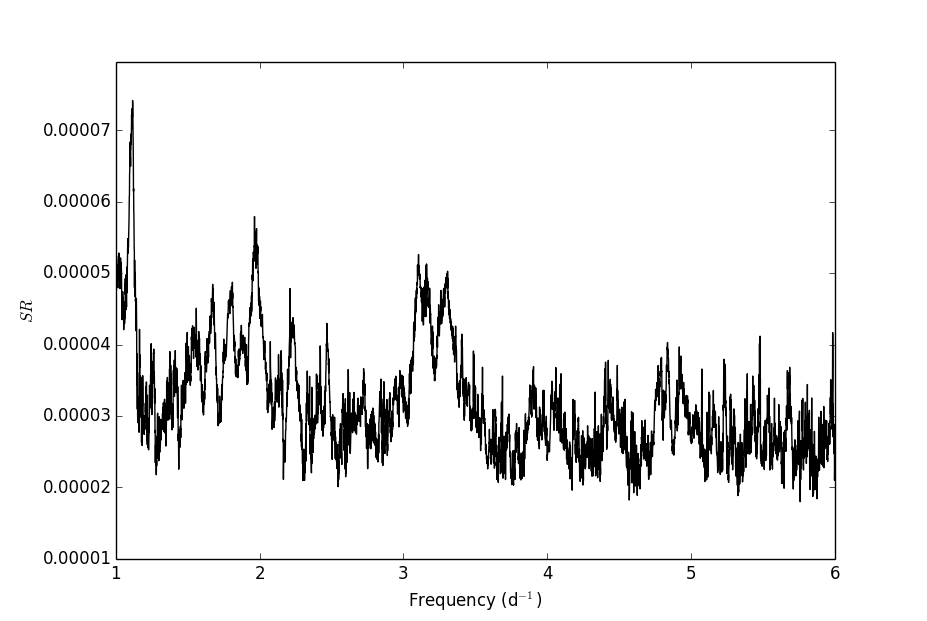}
\caption{Signal Residue as a function of orbital frequency for our 100 hour $Spitzer$ program. We subtract the best fitting transit model for GJ 1132b and use the box least square algorithm from \citet{BLS_algorithm} to compute the best-fitting box signal at each orbital period. We find no significant signals at short periods (P $<$ 50 hours). The broad peak located near P = 0.3 days (frequency = 3.3) is associated with the large positive flux deviation at 2457504.5 BJD and additional positive outliers in the data set. We find no significant periodic negative flux deviation signal in our dataset and exclude transits approximately three quarters the size of Mars at periods of 50 hours or less.
}
\label{BLS_figure}
\end{singlespace}
\end{figure}

Transiting bodies with orbital periods of 50 hours or less would show multiple transits during the span of our observation. With multiple transit measurements, we can increase our sensitivity to small bodies, provided we can search over the required period-space in order to coherently stack the transits together. In order to search for the presence of transiting bodies on ultra short periods, we first subtract the best-fitting transits of GJ 1132b from our $Spitzer$ dataset. We searched for periodic signals using the box least squares (BLS) method described by \citet{BLS_algorithm}. We plot the signal-residue as a function of orbital frequency in Figure \ref{BLS_figure}. The broad signal located at approximately at a period of 0.3 days is associated with positive flux outliers in the our data and not possible transit signals. The best fit ``depth" for this signal is -0.0002 (i.e. a brightening) of the total flux. We can exclude bodies the size of 0.74 times the size of Mars in orbital periods of 50 hours or less. We note that this limit is larger than $2^{=0.5}$ times the limit from the single transit case due to red noise in our $Spitzer$ light curve at large time scales.

\subsection{Limits on additional bodies from transit timing variations}
We see no significant deviations from a linear ephemeris from any of the measured central transit times. Our observations span 259 epochs, or 422 days. Only one observation is more than 5 minutes deviant from a linear ephemeris, and this observation has an error bar of 2.7 minutes, due to the relatively poorer weather conditions during this observation. Transit timing variations are largest (and most detectable) when the perturbing body is near a first-order mean-motion resonance \citep{Holman_TTV}, although perturbations from bodies in a second-order resonance are also detectable \citep{Deck_TTV_1}. Planets in retrograde orbits relative to the planet whose transit times are being measured also show diminished amplitudes, further restricting the sensitivity of a TTV analysis to placing upper limits on additional planetary bodies in a system \citep{TTV_Payne}. We see no evidence for any transit timing variations over this timescale, and therefore conclude that there is unlikely to be any bodies of significant mass near mean-motion resonances with GJ 1132b. 

\subsection{The dimming event at BJD 57503.69}
We see one significant outlier signal immediately after the large gap in our data after the $Spitzer$ data downlink. The depth of this signal is 0.022\%, and lasts for the first 3.5 hours after data collection resumes. This signal does not correlate with instrumental systematics such as pointing stability, voltage or current applied to the heater, or any of the temperature measurements on the $Spitzer$ spacecraft during these observations. Similar signals in long stares of exoplanet hosts, like GJ 1214, have not shown similar signals after data downlink events \citep{GJ1214_point_and_stare}. If this signal is real, it must begin at some point during the data downlink, and so 3.5 hours is the minimum transit duration for this signal. For a transit crossing the equator of the star, this corresponds to an orbital velocity of approximately 22 km/s and an orbital period of nearly 180 days. We note that if this signal is real that this body cannot be coplanar with GJ 1132b. Considering this extremely long duration, the unlikely chance that a body in an orbit with a period of 180 days would transit during a 100 hours observation window, the unlikely chance that if such a body existed that it would transit, and that this signal is coincident after a $Spitzer$ repointing after data downlink, we believe that this signal is not real and is likely to be due to a spacecraft systematic associated with the data downlink and repointing.

\subsection{Limits on the secondary eclipse of GJ 1132b}
\begin{figure}
\begin{singlespace}
\centering
\includegraphics[width=1.0\linewidth]{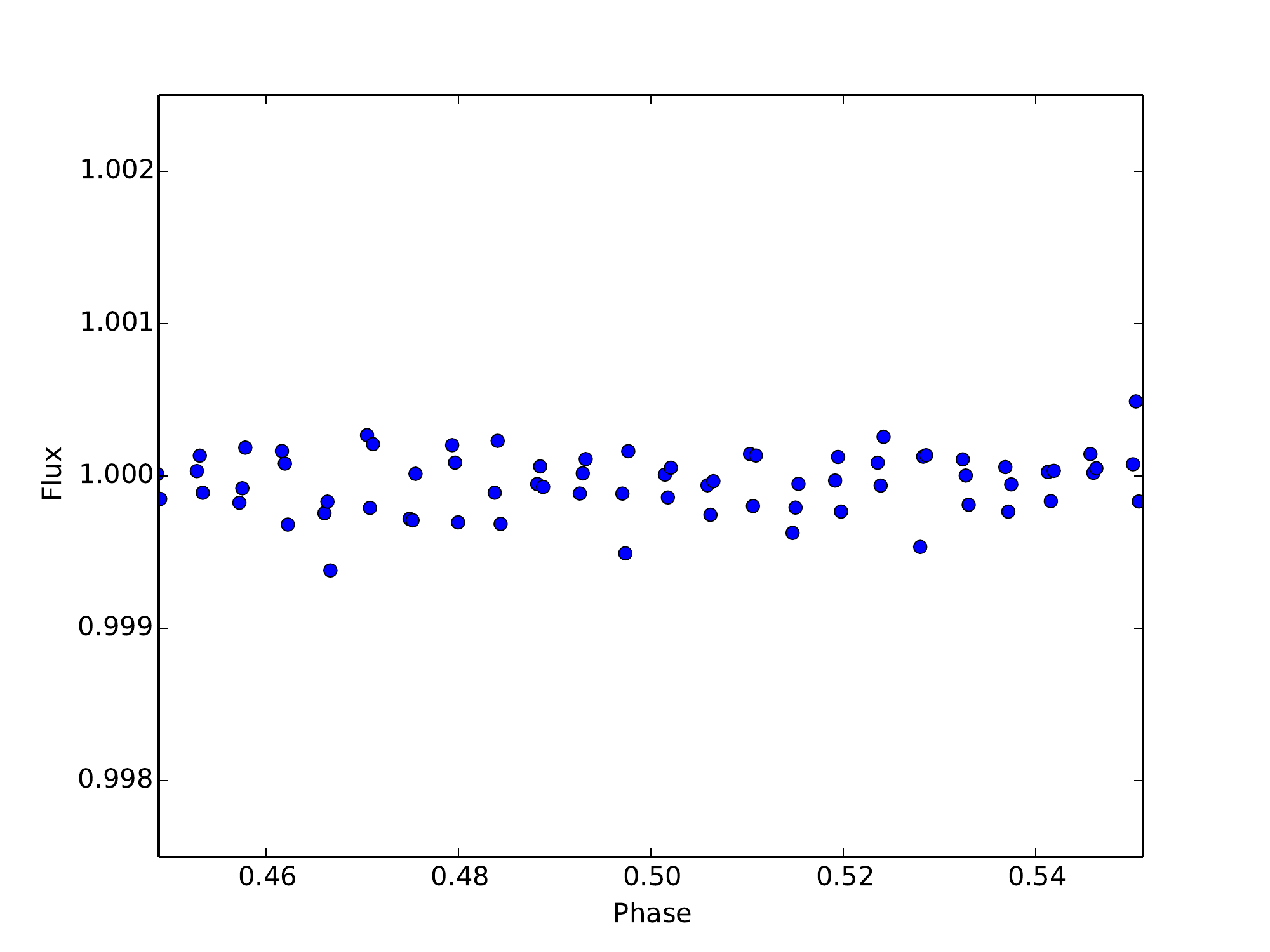}
\caption{$Spitzer$ observations phase folded to the ephemeris of GJ 1132b and binned on 10 minute timescales. We find no evidence for a secondary eclipse of GJ 1132b, and rule out eclipse depths of 480 ppm at $3\sigma$ confidence. The expected secondary eclipse depth for GJ 1132b at the equilibrium temperature for its semi-major axis is approximately 8 ppm, and therefore these observations are not very constraining on the nature of GJ 1132b's atmosphere. In order to be detected in these observations, GJ 1132b would need to have a blackbody temperature of 700K. 
}
\label{secondary_eclipse}
\end{singlespace}
\end{figure}

Our observations also contain three observations of the time of secondary eclipse of GJ 1132b (assuming zero eccentricity). The thermal variation expected from GJ 1132b assuming zero albedo and a temperature equal to the equilibrium temperature is 8 ppm. This is much smaller than the sensitivity of our $Spitzer$ data. GJ 1132b would need an effective temperature of approximately 700K if emitting as a blackbody in order to be detectable in our data. However, we can rule out secondary eclipses from extended warm atmospheres with our $Spitzer$ observations. In Figure \ref{secondary_eclipse}, we show our $Spitzer$ lightcurve phase folded to GJ 1132b's ephemeris and binned on 10 minute time scales. We find a $3 \sigma$ upper limit of 480 ppm for the secondary eclipse depth. This assumes that the eccentricity of GJ 1132 is zero, which has not yet been measured. We note that if GJ 1132b was eccentric, that our search for additional transiting bodies would also be sensitive to secondary eclipses from GJ 1132b regardless of GJ 1132b's eccentricity. Since we see no evidence for any additional transit signatures, and cover the full phase of GJ 1132b's orbit, this upper limit for secondary eclipses is robust to assumptions about GJ 1132b's eccentricity.

\subsection{Expectations for the GJ 1132 system from the $Kepler$ Statistics}
In this work we have found no evidence for additional transiting bodies with periods of 100 hours or less. We further find no evidence for transit timing variations from bodies in mean-motion resonance with GJ 1132b. However, the $Kepler$ dichotomy suggests that small rocky planets around M dwarfs (like GJ 1132b) are likely to host additional coplanar planets with periods less than $10$ days, while single planetary systems (like GJ 436b and GJ 1214b) are more likely to be larger with significant gaseous envelopes. If GJ 1132 hosts additional planetary bodies, they must either be \emph{a)} smaller than the size of Mars such that any transits by these bodies would be undetected in our data, \emph{b)} mutually inclined with GJ 1132b such that they do not show a transiting geometry when viewed from the Earth, or \emph{c)}, at orbital periods longer than 100 hours such that they did not transit during the timespan of our observations. Detecting bodies smaller than GJ 1132b, through either future transit measurements or RV measurements will be difficult, as the signal size is small. However, planets at longer period orbits can potentially be detected by future transit observations by, for example, TESS (if they transit), or through a sustained RV observational campaign if the mass of the planet is large enough to be detectable.

\acknowledgments
This work is based in part on observations made with the Spitzer Space Telescope, which is operated by the Jet Propulsion Laboratory, California Institute of Technology under a contract with NASA. Support for this work was provided by NASA through an award issued by JPL/Caltech. JAD acknowledges Ben Montet for helpful conversations regarding working with $Spitzer$ data. The MEarth Team gratefully acknowledges funding from the David and Lucille Packard Fellowship for Science and Engineering (awarded to D.C.). This material is based upon work supported by the National Science Foundation under grants AST-0807690, AST-1109468, AST-1004488 (Alan T. Waterman Award), and AST-1616624. ERN is supported by an NSF Astronomy and Astrophysics Postdoctoral Fellowship. This publication was made possible through the support of a grant from the John Templeton Foundation. The opinions expressed in this publication are those of the authors and do not necessarily reflect the views of the John Templeton Foundation. This research has made extensive use of NASA's Astrophysics Data System (ADS), and the SIMBAD database, operated at CDS, Strasbourg, France.


\clearpage

\bibliography{bibliography}

\clearpage


\begin{singlespace}
\begin{table}
\begin{center}
\caption{Photometry of GJ 1132}
\label{Photometry_table}
\begin{tabular}{crrrrr} 
\tableline\tableline
\centering
Time (BJD) & Flux & Error & Instrument \\
\tableline
\tableline 
2457332.742537 & 0.9917 & 0.0053 & MEarth \\
2457332.743049 & 1.0053 & 0.0052 & MEarth \\
2457332.743572 & 0.9997 & 0.0052 & MEarth \\
2457332.744091 & 1.0064 & 0.0052 & MEarth \\
2457332.744603 & 0.9937 & 0.0051 & MEarth \\
... & ... & ... & ... \\
\tableline
\end{tabular}
\end{center}
\end{table}
\end{singlespace}

\clearpage
\newpage
\begin{singlespace}
\begin{table}
\begin{center}
\caption{Transit Model Parameters}
\label{model_parameters}
\begin{tabular}{crrrr} 
\tableline\tableline
\centering
Parameter & Description \\
\tableline
\tableline 
$\frac{a}{R_*}$ & Ratio of orbital semi-major axis and stellar radius \\
$i$ & Orbital inclination angle in the plane of the sky (degrees) \\
$\frac{R_p}{R_*}$ & Ratio of planetary radius and stellar radius \\
$P$ & Orbital Period (days) \\
$T_0$ & Transit epoch (BJD) \\
$a_{4.5}$ & Quadratic limb darkening law coefficient 1 in $Spitzer$ Channel 2 \\
$b_{4.5}$ & Quadratic limb darkening law coefficient 2 in $Spitzer$ Channel 2 \\
$a_{\textrm{MEarth}}$ & Quadratic limb darkening law coefficient 1 in the MEarth-South bandpass \\
$b_{\textrm{MEarth}}$ & Quadratic limb darkening law coefficient 2 in the MEarth-South bandpass \\
\tableline
\end{tabular}
\end{center}
\end{table}
\end{singlespace}

\begin{singlespace}
\begin{table}
\begin{center}
\caption{System Parameters}
\label{model_fit}
\begin{tabular}{crrrrrr} 
\tableline\tableline
\centering
Parameter & Value & Source  \\
\tableline
\tableline 
$\frac{a}{R_*}$ & $16.54^{+0.63}_{-0.71}$ & This work\\
$i$ (degrees) & $88.68^{+0.40}_{-0.33}$ & This work \\
$\frac{R_p}{R_*}$ ($Spitzer$) & $0.0492 \pm 0.0008$ & This work \\
$\frac{R_p}{R_*}$ (MEarth) & $0.0455 \pm 0.0006$ & This work \\
$P$ (days) & $1.6289246^{+0.0000024}_{-0.0000030}$ & This work \\
$T_0$ (BJD) & $2457184.55804^{+0.00054}_{-0.00039}$ & This work \\
$a_{4.5}$ & $0.313^{+0.0041}_{-0.0042}$ & This work \\
$b_{4.5}$ & $0.154^{+0.022}_{-0.018}$ & This work \\
$a_{\textrm{MEarth}}$ & $0.215^{+0.016}_{-0.027}$ & This work \\
$b_{\textrm{MEarth}}$ & $0.407^{+0.049}_{-0.064}$ & This work \\
$R_*$ ($R_\odot$) & $0.2105^{+0.0102}_{-0.0085}$ & This work \\
$\rho_*$ ($\rho_\odot$) & $19.4^{+2.6}_{-2.5}$ & This work \\
$M_*$ ($M_\odot$) & $0.181 \pm 0.019$ & \citet{GJ1132_discovery} \\
$M_p$ ($M_\oplus$) & $1.62 \pm 0.55$ & \citet{GJ1132_discovery} \\
\tableline
\end{tabular}
\end{center}
\end{table}
\end{singlespace}

\newpage
\begin{singlespace}
\begin{table}
\begin{center}
\caption{Individual Transit Times}
\label{timeing_table}
\begin{tabular}{crrrr} 
\tableline\tableline
\centering
Epoch & Transit Time (BJD) & Error & & Instrument \\
\tableline
\tableline 
-19 & 2457153.6079 & 0.0024 & MEarth \\
-11 & 2457166.63969 & 0.00046 & MEarth \\
0 & 2457184.55789 & 0.00031 & MEarth + TRAPPIST + PISCO \\
11 & 2457202.47611 & 0.00034 & MEarth \\
91 & 2457332.7943 & 0.0019 & MEarth \\
99 & 2457345.81951 & 0.00089 & MEarth \\
110 & 2457363.7397 & 0.0010 & MEarth \\
118 & 2457376.7712 & 0.0013 & MEarth \\
121 & 2457381.6588 & 0.0012 & MEarth \\
134 & 2457402.8332 & 0.0017 & MEarth \\
137 & 2457407.72076 & 0.00090 & MEarth \\
145 & 2457420.7519 & 0.0011 & MEarth \\
148 & 2457425.63902 & 0.00096 & MEarth \\
153 & 2457433.7844 & 0.0011 & MEarth \\
156 & 2457438.6698 & 0.0013 & MEarth \\
159 & 2457443.55700 & 0.00088 & MEarth \\
161 & 2457446.8162 & 0.0016 & MEarth \\
164 & 2457451.7030 & 0.0018 & MEarth \\
167 & 2457456.5872 & 0.0012 & MEarth \\
178 & 2457474.5080 & 0.0012 & MEarth \\
180 & 2457477.76452 & 0.00081 & MEarth \\
186 & 2457487.5385 & 0.0017 & MEarth \\
191 & 2457495.6831 & 0.0011 & MEarth \\
194 & 2457500.5695 & 0.0013 & MEarth \\
196 & 2457503.82680 & 0.00048 & $Spitzer$ \\
197 & 2457505.45606 & 0.00051 & $Spitzer$ \\
240 & 2457575.49988 & 0.00081 & MEarth \\
\tableline
\end{tabular}
\end{center}
\end{table}
\end{singlespace}

\newpage

\newpage

\newpage

\newpage

\newpage

\newpage

\newpage

\newpage

\end{document}